# Locality-Aware Process Scheduling for Embedded MPSoCs*

*Mahmut Kandemir and Guilin Chen*
*Computer Science and Engineering Department*
*The Pennsylvania State University, University Park, PA 16802, USA*
*{kandemir, guilchen}@cse.psu.edu*

**Abstract**

Utilizing on-chip caches in embedded multiprocessor-system-on-a-chip (MPSoC) based systems is critical from both performance and power perspectives. While most of the prior work that targets at optimizing cache behavior are performed at hardware and compilation levels, operating system (OS) can also play major role as it sees the global access pattern information across applications. This paper proposes a cache-conscious OS process scheduling strategy based on data reuse. The proposed scheduler implements two complementary approaches. First, the processes that do not share any data between them are scheduled at different cores if it is possible to do so. Second, the processes that could not be executed at the same time (due to dependences) but share data among each other are mapped to the same processor core so that they share the cache contents. Our experimental results using this new data locality aware OS scheduling strategy are promising, and show significant improvements in task completion times.

## 1. Introduction

Today's embedded applications are very different from those in the past, in terms of both application complexity and dataset sizes. Consequently, it is not feasible any more to meet demands of embedded applications by using single core based systems. Instead, we observe a growing trend towards employing multiprocessor-system-on-a-chip (MPSoC) type of architectures, where multiple processor cores reside on the same chip, and share data through on-chip memory and/or on-chip communication network. While hardware support for embedded MPSoCs is extremely important (and it is in fact an active research area), we believe that task mapping and scheduling at the OS level is an area that can significantly affect overall performance.

*This work is supported in part by NSF Career Award #0093082.

One of the critical issues in embedded MPSoCs is on-chip memory performance. This is because if on-chip memory is not utilized appropriately, one can incur a significant number of off-chip references (accesses), which can be very expensive from both performance and power perspectives. Since it is possible to execute multiple tasks (each with multiple processes) on a given MPSoC, the behavior of on-chip memory components of an MPSoC is not only determined by data access patterns of individual processes but also through the interaction among access patterns of different processes. Consequently, to maximize utilization of on-chip memory, one needs to take into account this interaction as well.

To illustrate the problem, let us consider the following scenario. Suppose that we have 8 processor cores in our MPSoC, and each core can execute one process at any given time. Let us assume that a task called Task1 is scheduled across these cores and another task called Task2 is scheduled next across the same cores. Assume further that the scheduler schedules the processes of these tasks iteratively in an interleaved manner (i.e., Task1, Task2, Task1, Task2, Task1,…). In this case, the data brought by the processes of Task1 into on-chip caches in the MPSoC can be displaced by those brought by the processes of Task2, and vice versa. As a consequence, the next time Task1 (Task 2) is scheduled, its processes may experience a large number of cache misses, degrading overall performance.

There are at least two potential complementary solutions to this problem. First, the processes that do not share any data between them should be scheduled at different cores if it is possible to do so (in an attempt to minimize the conflicts between them). The second potential solution is that the processes that could not be executed at the same time (e.g., due to data and/or control dependences) but share data among each other should be mapped to the same processor core so that they share cache contents. Focusing on array-intensive embedded applications from image and video processing domains and a cache-based embedded MPSoC architecture, this paper presents an OS scheduling strategy that incorporates these two solutions. The basic idea behind our scheduler is to capture inter-process data sharing and utilize this

```
for(i1=0; i1 < 8; i1++)                     Task[i1] (i1 = 0, 1, 2, …, 7):
  for(i2=0; i2 < 3000; i2++)          ⟹     for(i2 =0; i2 < 3000; i2 ++)
    B[i1] += A[i1*1000 + i2][5]                B[i1] += A[i1*1000 + i2][5]

              Prog1

for(i1=0; i1 < 8; i1++)                     Task[i1] (i1 = 0, 1, 2, …, 7):
  for(i2=0; i2 < 3000; i2++)          ⟹     for(i2 =0; i2 < 3000; i2 ++)
    B[i1] += D[i1*1000 + i2][5]                B[i1] += D[i1*1000 + i2][5]

              Prog2
```

**Figure 1. Examples of parallelizing tasks over eight processes.**

|  | $P_0$ | $P_1$ | $P_2$ | $P_3$ | $P_4$ | $P_5$ | $P_6$ | $P_7$ |
|---|---|---|---|---|---|---|---|---|
| $P_0$ |  | 2000 | 1000 | 0 | 0 | 0 | 0 | 0 |
| $P_1$ | 2000 |  | 2000 | 1000 | 0 | 0 | 0 | 0 |
| $P_2$ | 1000 | 2000 |  | 2000 | 1000 | 0 | 0 | 0 |
| $P_3$ | 0 | 1000 | 2000 |  | 2000 | 1000 | 0 | 0 |
| $P_4$ | 0 | 0 | 1000 | 2000 |  | 2000 | 1000 | 0 |
| $P_5$ | 0 | 0 | 0 | 1000 | 2000 |  | 2000 | 1000 |
| $P_6$ | 0 | 0 | 0 | 0 | 1000 | 2000 |  | 2000 |
| $P_7$ | 0 | 0 | 0 | 0 | 0 | 1000 | 2000 |  |

(a)

| Core | 0 | 1 | 2 | 3 |
|---|---|---|---|---|
| $T_1$ | $P_0$ | $P_2$ | $P_4$ | $P_6$ |
| $T_2$ | $P_1$ | $P_3$ | $P_5$ | $P_7$ |

(b)

| Core | 0 | 1 | 2 | 3 |
|---|---|---|---|---|
| $T_1$ | $P_0$ | $P_2$ | $P_4$ | $P_6$ |
| $T_2$ | $P_3$ | $P_5$ | $P_7$ | $P_1$ |

(c)

**Figure 2. (a) Data sharings between different processes. (b) Mapping and scheduling of the processes with good data reuse. (c) Mapping and scheduling of the processes with poor data reuse. We assume four processor cores.**

information in scheduling processes across multiple cores. We also propose a data re-mapping strategy to reduce the number of conflict misses between the processes that do not share data.

We implemented our approach within a simulation environment and tested its effectiveness using a set of tasks (applications) under different execution scenarios. Our experimental evaluation indicates that the proposed data-aware scheduling strategy is very successful in practice. We also observe that our savings are consistent across several simulation parameters.

The rest of this paper is organized as follows. Section 2 explains how we capture data sharings between different processes. Section 3 gives the details of our OS scheduling scheme. Section 4 presents an experimental evaluation of the proposed scheduling strategy. Section 5 discusses the related work in this area, and Section 6 concludes the paper and outlines our planned future efforts along this direction.

## 2. Capturing Inter-Process Data Sharing

To represent sets of data elements shared among processes, we use Presburger arithmetic. Our approach can be best explained using an example. Consider the two program fragments, Prog1 and Prog2, shown in Figure 1, each representing a task. The iteration space of the first fragment can be represented using the following set:

$$IS_1 = \{[i_1,i_2]: 0 \le i_1 < 8 \ \&\& \ 0 \le i_2 < 3000\}.$$

Suppose that this task is parallelized in such a fashion, over 8 cores, that each process receives a set of successive loop iterations. Therefore, the iteration set of process k (where $0 \le k \le 7$) can be represented as:

$$IS_{1,k} = \{[i_1,i_2]: i_1 = k \ \&\& \ 0 \le i_2 < 3000\}.$$

Consequently, the set of data elements accessed by the $k^{th}$ process is:

$$DS_{1,k} = \{[d_1,d_2]: d_1=i_1*1000+i_2 \ \&\& \ d_2=5 \ \&\& \ [i_1,i_2] \in IS_{1,k}\}.$$

Using $DS_{1,k}$ and $DS_{1,p}$, one can express the set of elements shared by processes k and p as follows:

$$SS_{1,k,p} = DS_{1,k} \cap DS_{1,p}$$

Similar sets can be written for every processor pairs. Note that the impact of these data sharings depends on the order in which these processes are scheduled (executed). If the processes are scheduled at the same time (i.e., concurrently), these sharings can cause data being duplicated across multiple on-chip caches. In this case, there is nothing much that can be done to take advantage of these data sharing from the scheduler's perspective. On the other hand, if k and p are scheduled one after another (e.g., due to lack of sufficient resources to execute them in parallel), one would prefer that they execute on the same processor core if they share data, or in different cores otherwise.

For our running example, Figure 2(a) illustrates the sharings between the different processor cores, assuming that we have four cores to execute the processes. Each cell (k,p) in this table gives the amount of data shared between process k and process p. Now, let us assume that processes 0, 2, 4, and 6 are scheduled at time quantum 1 ($T_1$) and the remaining four are scheduled at time quantum 2 ($T_2$). From the data sharing viewpoint, the best process-to-core mapping is as shown in Figure 2(b). Notice that, in this mapping, the processor pairs share significant amount of data, which should normally result in a very good on-chip cache behavior. As comparison, the alternate mapping depicted in Figure 2(c) for example would not lead to any data sharing, and thus, is not preferable from the cache utilization perspective. Let us now focus on Prog2 and the data sharing among its processes. Note that formulations similar to those of Prog1 can be developed for this task (Prog2) as well. We observe that the processes of Prog2 accesses array D that is not used in the processes of Prog1. Depending on how the arrays A and D are laid in the main memory, there can be cache conflicts (due to limited associativity). Consequently, each time a process (of Prog1 or Prog2) is scheduled, it would have only little opportunity (if any at all) for data reuse. In the next section, we give the details of our task scheduling

```
Input:
   PS: the set of processes;    X: the number of cores
   M[1..X][1..X]: sharing matrix (see Figure 2(a))
Output:
   Process scheduling of each core
procedure schedule(PS, X) {
   IN = {p | p ∈ PS and p does not depend on any process};
   PS = PS – IN;
   while(|IN| > X) {
      select p ∈ IN such that Σ_{q ∈ IN} M[p][q] is minimized;
      IN = IN - {p}; PS = PS + {p}
   }
   schedule the processes in IN on the cores
   while(PS is not empty) {
      for k = 1 to X {
         p = the previous scheduled process on core[k];
         select q ∈ PS
            such that q does not depend on any unscheduled
             process and M[p][q] is maximized;
         Schedule q on core[k];
         PS = PS – {q}
      }
   }
}
```

**Figure 3. Locality-aware process scheduling algorithm**

strategy that minimizes such conflicts and optimizes the reuse of on-chip data as much as possible.

# 3. Details of Data Reuse Oriented Scheduling

In our framework, each task is represented using a process graph (denoted PG). Each node in the process graph of task i corresponds to a process j, and is denoted using $P_{i,j}$. A directed edge from $P_{i,j}$ to $P_{i,k}$ indicates that the latter is dependent on the former. That is, $P_{i,k}$ can execute only when $P_{i,j}$ finishes its execution. Note that, depending on how the tasks in the system are constructed, there may also be dependences between processes that belong to different tasks (e.g., from $P_{i,k}$ to $P_{i',l}$ where $i \neq i'$). Therefore, we can talk about an extended process graph (denoted EPG) that involves these inter-task dependences as well (in addition to intra-task dependences). When we are discussing an EPG, we can assume that each process has a unique id, and thus we remove its task id when there is no confusion. Our scheduling problem can then be defined as one of scheduling a given EPG such that on-chip (L1) data reuse is maximized to the greatest extent possible. In the following paragraphs, we discuss the details of our scheduling approach.

Our approach to the scheduling problem is a greedy heuristic that is based upon data sharing between the processes. A formal sketch of our algorithm is given in Figure 3. In the initialization step of this algorithm, we identify the independent processes (i.e., the processes without an incoming dependence edge in the EPG) that can be mapped to the cores. These processes are the candidates for the first round of scheduling. However, if

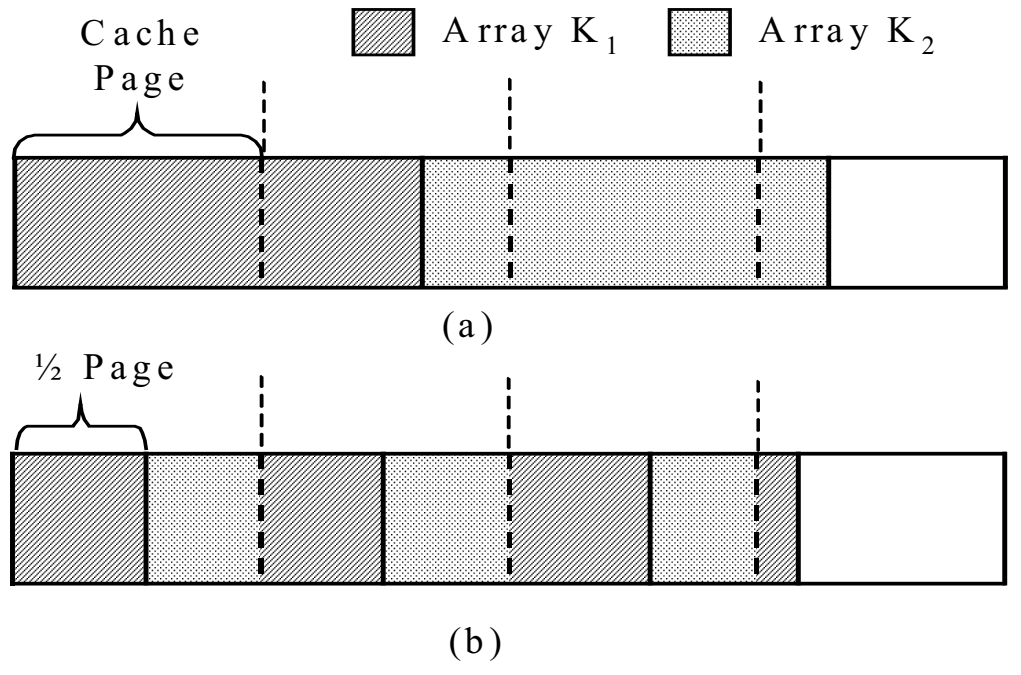

**Figure 4. An example data mapping to reduce conflict misses.**

```
Input:
   A_1, A_2, …, A_n: arrays that are accessed by these processes;
   M[1..n][1..n]: conflict matrix
   C: cache page size;
   T: threshold
Output:
   optimized layout of A_1, A_2, …, A_n;
procedure re-layout() {
   select (x, y) such that M[i][j] is maximized;
   while(M[x][y] > T) {
      M[x][y] = 0;
      if(A_x and A_y that are accessed by the same process,
         or respectively accessed by a pair of processes that
         are scheduled successively on the same core) {
            if(A_x has been re-layouted) {
               re-layout A_y to avoid conflict with A_x;
            } else if(A_y has been re-layouted){
               re-layout A_x to avoid conflict with A_y;
            } else {
               re-layout both A_x and A_y to avoid conflict;
            }
      }
      select (x, y) such that M[x][y] is maximized
                   and that A_x or A_y has not been re-layouted;
   }
}
```

**Figure 5. The algorithm for selection of the arrays to be-layouted**

we have a larger number of candidates than the number of available cores, we need to make a selection from these candidates such that the data sharing between them is kept at the minimum. This makes sense since these processes are independent of each other and they will be mapped onto different cores. Our greedy scheduling algorithm iteratively removes the candidates that have the maximum data sharing with the other candidates until the number of the remaining candidates is equal to the number of the cores.

After the initialization step, the main body of the algorithm is executed. At each iteration of the while loop shown, we select for each core the next process to be scheduled on it. Our selection criterion this time is to maximize the data reuse between the selected process and the one that was scheduled on the same core in the previous step (scheduling quantum). Mathematically, supposing that process i was the current process scheduled on a particular core, the next process (denoted j) to be scheduled on the same core is selected such that:

$$|SS_{i,j}| \geq |SS_{i,k}|$$

for any k≠j. This approach clearly tries to maximize the reuse of on-chip data for any given core. It should be noticed, however, that since the cores are processed (by the algorithm) in a specific order, it is possible that this algorithm does not generate the best results (from data reuse perspective) in all cases.

**Table 1. Applications used in this study.**

| Applications (Task) | Brief Description |
|---|---|
| Med-Im04 | medical image reconstruction |
| MxM | triple matrix multiplication |
| Radar | radar imaging |
| Shape | pattern recognition and shape analysis |
| Track | visual tracking control |
| Usonic | feature-based object recognition |

An important point to note here is that, while the algorithm shown in Figure 3 tries to maximize data reuse, it is still possible for two processes without any data sharing to be scheduled on the same processor core successively. When this happens, data locality can be very poor depending on the number of cache conflicts. Next, we discuss this issue in detail.

The purpose of data mapping is to assign elements of each array of an application to memory locations. However, our goal in this mapping is to minimize the cache conflicts between the array elements manipulated by different processes that are scheduled on the same core. A cache conflict is said to occur if two array elements, $A[d_1, d_2]$ and $B[d_3, d_4]$, map to the same cache line. This can be expressed more formally as:

$$map(addr(A[d_1, d_2])) = map(addr(B[d_3, d_4])),$$

where *addrs*(.) is a function that maps each array element to a main memory address, and *map*(.) is a function that maps a given memory address to a cache line. The function *addr*(.) is determined by the specification of the programming language or the implementation of compiler. The *map*(.) is determined by cache parameters (e.g., cache size, associativity, and block size). For a multiple-way associative cache, we cannot statically determine which cache line a memory address may be mapped to. In this case, *map*() maps each address to a set of cache lines.

**Table 2. Default simulation parameters used in this study.**

| Parameter | Value |
|---|---|
| Number of processors | 8 |
| Data/instruction cache per processor | 8KB, 2-way |
| Cache access latency | 2 cycle |
| Off-chip memory access latency | 75 cycles |
| Processor speed | 200 MHz |

The cache line into which an array element can be loaded is determined by both *addr*(.) and *map*(.). We usually do not have the flexibility to change *map*(.) since it is determined by the hardware. On the other hand, a compiler can change *addr*(.) by changing the layouts of the arrays to reduce the conflicts. Figure 4 illustrates how we re-layout two arrays $K_1$ and $K_2$ to avoid the conflicts between them. Let us assume that we have two processes, $p_1$ and $p_2$, and that process $p_2$ is scheduled right after $p_1$. Assume further that process $p_1$ iteratively accesses each element of both arrays $K_1$ and $K_2$, and process $p_2$ accesses only $K_2$. Figure 4(a) shows the original memory layout of the two arrays. Using such layout, the elements of $K_1$ and $K_2$ may be mapped to the same cache line, which can create severe conflicts. Figure 4(b) shows our optimized layouts for arrays $K_1$ and $K_2$. Specifically, we divide each array into a set of chunks. The size of each chunk is equal a half of one cache page[1]. Further, the chunks of arrays $K_1$ and $K_2$ interleave with each other. Using such a layout, the elements of $K_1$ cannot be mapped to the same cache lines to which the elements of $K_2$ are mapped. The following formula shows how we calculate the new address of each array element:

$$addr'(A[x, y]) = 2addr(A[x, y]) - addr(A[x, y]) \bmod (C/2) + b$$

where *addr'*(.) and *addr*(.) are the optimized and the original main memory address mappings; *C* is the size of cache page; *b* is either 0 or *C*/2. Note that two arrays using different values of b can never conflict with each other. For example, in Figure 4(b), $K_1$ is mapped using $b$=0 while $K_2$ is mapped using $b = C/2$. By re-layouting $K_1$ and $K_2$, we avoid the conflicts between them, which is beneficial to the performance of $p_1$. Further, when $p_2$ is scheduled, many elements of $K_2$ are already in the cache, which also improves the performance of $p_2$. In Figure 5, we give the sketch of a greedy heuristic algorithm that selects the arrays for re-layouting. Specifically, in the while loop of the algorithm, we iteratively select the pair of arrays with maximum number conflicts with each other. If these arrays are accessed by the same process or respectively accessed by a pair of processes that are scheduled successively on the same core, we try to re-layout them to avoid cache conflict. It should be noted that, if both the arrays have been re-layouted in a previous execution of the while loop, we do not attempt to re-layout either of them since their layouts have been determined to avoid the conflicts whose conflict number is larger than the current one. The algorithm terminates when we cannot find a pair of arrays between which the conflict number is larger than the given threshold T. In our experiments, we set T to the average number of conflicts across all pairs of arrays.

[1] Size of a cache page = cache size / cache associativity.

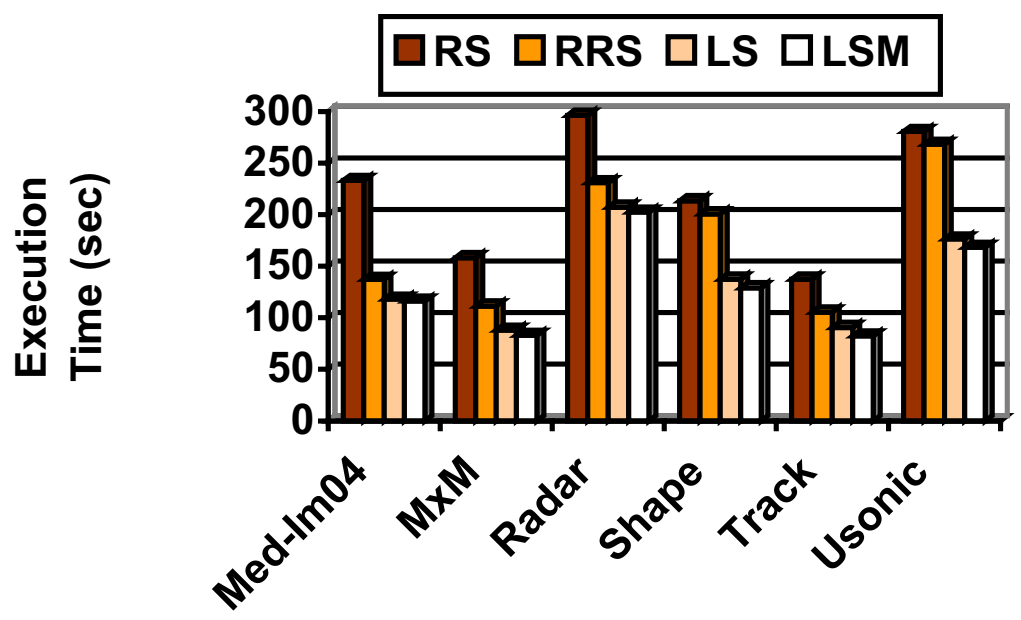

**Figure 6. Execution times when applications are executed in isolation.**

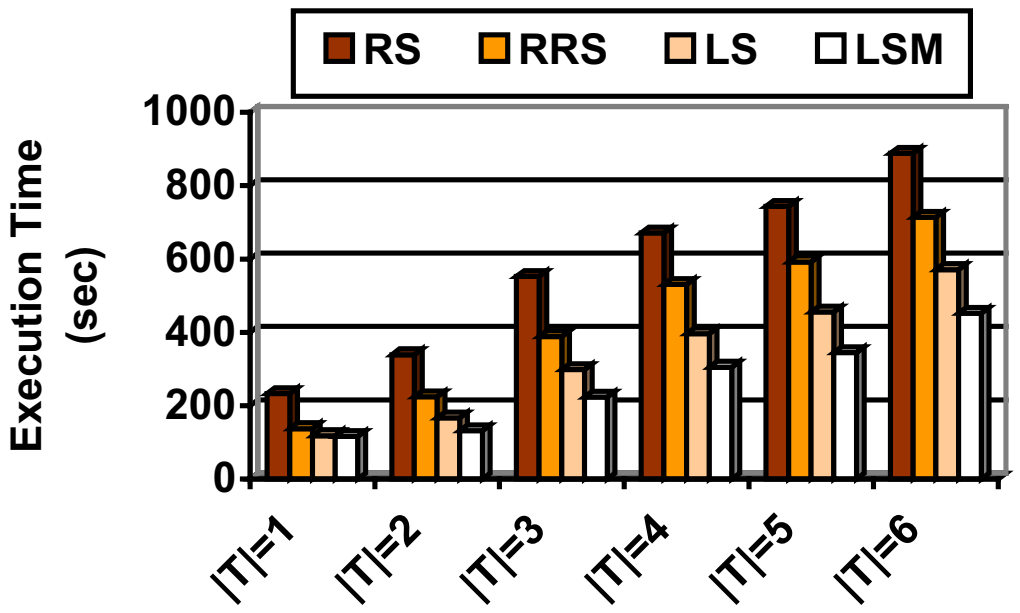

**Figure 7. Execution times when applications are executed concurrently.**

## 4. Experimental Evaluation

In our experiments, we used six applications (tasks). A common characteristic of these applications is that they are all array-based embedded codes from the domain of image/video processing. The brief descriptions of these applications are given in Table 1. The numbers of processes of these benchmarks (tasks) vary between 9 and 37. We evaluate our scheduling approach using these applications in two different ways. First, we consider each application in isolation, i.e., we perform scheduling, assuming that only the processes of the application in question are running on the MPSoC. Then, we schedule processes from multiple applications when they execute concurrently.

We performed our experiments using Simics tool [9]. Simics runs operating systems, device drivers, and firmware in addition to the application code. This means that the applications executed on Simics will behave as they would on a real system, while executing just the user-level application code will not reveal all properties of the system. In this work, we experimented with different schedulers implemented on top of Simics. The default simulation parameters used in the experiments are given in Table 2.

We tested four different process scheduling strategies:

1. *Random Scheduling (RS):* In this strategy, each process is assigned to an available core randomly without any concern for data reuse. Once scheduled, each process runs to completion (unless it needs to synchronize with another process).

2. *Round-Robin Scheduling (RRS):* This is a preemptive FCFS (first-come-first served) scheduling. Basically, we maintain a ready queue for processes (as FIFO). New processes are added to the tail of the queue, and the scheduler selects the first process from the ready queue, sets a timer, and schedules it. When the timer is off, the process relinquishes the core voluntarily, and the next one in the queue is scheduled. Note that all cores take their processes from a common ready queue.

3. *Locality-Aware Scheduling without Data Mapping (LS):* This is the scheduling strategy discussed in this paper, except that it does not include the data mapping part discussed in Section 3.

4. *Locality-Aware Scheduling with Data Mapping (LSM):* This is similar to LS except that it also includes the data mapping phase explained in Section 3.

The graph in Figure 6 gives the execution times of our applications when they are scheduled in isolation. One can make two major observations from these results. First, our locality-aware scheduling strategy generates much better results than both RS and RRS. This is because, in this isolated execution case, processes executing at a given time are from the same application. As a consequence, they share a lot of data, i.e., data reuse among them is high, and this makes cache locality behavior extremely important. Since neither RS nor RRS does anything special to exploit data locality, they perform poorly compared to both LS and LSM. The second observation is that the difference between LS and LSM it not too great. This is mainly because of the fact that the data sharing between processes dominates the cache conflicts (as the processes are from the same application).

Figure 7 shows the execution times under different workloads. On the x-axis, |T| gives the number of tasks (applications) running on the MPSoC. Therefore, as we move along this axis from left to right we keep introducing more tasks to the system (i.e., generate more pressure). More specifically, when |T|=1, we have only

(processes of) Med-Im04 running on the MPSoC; when |T|=2, we introduce MxM in addition to Med-Im04, and they execute concurrently; and so on. Therefore, the execution time in this case means the overall completion time of the applications executing concurrently. When we compare these results with those given in Figure 6, the most striking difference is that here there is a larger difference between LS and LSM. The main reason for this is that, since processes scheduled on the same core in this case can come from different applications (and since these applications do not share data among them), they can create too many conflict misses, most of which are eliminated by LSM but not LS.

## 5. Discussion of Related Work

The prior work on process scheduling in the embedded systems area include works targeting instruction and data caches. [10] and [6] present scheduling strategies for instruction caches that try to reduce the number of conflict misses. Li and Wolfe [7] propose and evaluate a model for estimating the performance of multiple processes sharing a cache. More recently Kadayif et al [3] and Kandemir et al [4] propose locality-conscious scheduling strategies for data caches. The main difference between our work and these two studies is that we focus on an MPSoC based architecture instead of a single processor based system. Riveral and Tseng [8] discuss techniques that minimize the number of conflict misses by transforming data layouts. Their work is focused on single processor environment. Our work is different from theirs in that the focus of our work is to minimize the cache misses among the data elements accessed by different processes in an embedded MPSoC based system. Carr et al [2] propose a cost model that computes both temporal and spatial reuse of cache lines to find desirable loop restructurings. Bershad et al [1] propose a scheme that uses a hardware-based Cache Miss Lookaside (CML) buffer to detect cache conflicts by recording and summarizing a history of cache misses. Whenever a large number of conflict misses is detected, a software policy (within the operating system) is invoked to remove the conflicts by dynamically re-mapping pages. Kharbutli et al [5] present an in-depth analysis on the performance degradation due to the pathological behavior of cache hashing functions in an architecture that uses cache rehashing to reduce conflict misses. Based on their analysis, they propose two hashing functions (prime modulo and prime displacement) that are resistant to pathological behavior and yet are able to eliminate the worst-case conflict behavior in the L2 cache.

## 6. Concluding Remarks and Future Research

Performance of applications executing on embedded MPSoCs depends strongly on their on-chip cache behavior. Significant gains are possible from both performance and power angles by being careful in scheduling processes on embedded cores in such a fashion that data reuse among successively scheduled processes (on the same core) is maximized. This paper presents an OS process scheduling strategy based on this idea, and evaluates its effectiveness using a suite of six embedded applications. Our results clearly emphasize the importance of cache-conscious process scheduling. This work can be extended in several directions. First, we plan to compare it to other OS scheduling strategies as well using our benchmarks. Second, we want to implement it within an embedded Linux environment and test it. Third, we want to explore the interactions between OS scheduling and compiler-driven techniques and identify the potential mechanisms through which these two can communicate.